\documentclass[aip,reprint,twocolumn,showpacs,floatfix,preprintnumbers,superscriptaddress,amsmath,amssymb]{revtex4-1}
\usepackage{amsmath}
\usepackage{amssymb}
\usepackage{graphicx}
\usepackage{epsfig}
\usepackage{epstopdf}
\usepackage{dcolumn}
\usepackage{bm}
\usepackage{array}
\usepackage{mathtools}
\usepackage[colorlinks,bookmarks=false,citecolor=blue,linkcolor=red,urlcolor=blue]{hyperref}

\begin{document}

\title{A simple atomic beam oven with a metal thermal break}
\author{Chetan Vishwakarma}
\affiliation{Department of Physics, Indian Institute of Science Education and Research, Pune 411008, Maharashtra, India}
\author{Jay Mangaonkar}\affiliation{Department of Physics, Indian Institute of Science Education and Research, Pune 411008, Maharashtra, India}
\author{Kushal Patel}
\affiliation{Department of Physics, Indian Institute of Science Education and Research, Pune 411008, Maharashtra, India}
\author{Gunjan Verma} 
\affiliation{Department of Physics, Indian Institute of Science Education and Research, Pune 411008, Maharashtra, India}
\author{Sumit Sarkar}
\affiliation{Department of Physics, Indian Institute of Science Education and Research, Pune 411008, Maharashtra, India}
\author{Umakant D. Rapol}
\email { umakant.rapol@iiserpune.ac.in}
\affiliation{Department of Physics, Indian Institute of Science Education and Research, Pune 411008, Maharashtra, India}
\affiliation {Center for energy sciences, Indian Institute of Science Education and Research, Pune 411008, Maharashtra, India}

\begin{abstract}

We report the design and construction of a simple, easy to machine high temperature oven for generating an atomic beam in laser cooling experiments. This design eliminates the problem of thermal isolation of the oven region from the rest of the vacuum system without using a glass or ceramic thermal break. This design simplifies the construction and operation of high temperature ovens for elements having low vapor pressure. We demonstrate the functionality of such a source for Strontium (Sr) atoms. We generate a high flux of Sr atoms for use in laser cooling and trapping experiments. 
The optimization of the design of the metal thermal break is done using a finite element analysis.

\end{abstract}

\maketitle

\section{Introduction}

Laser-cooled atomic samples provide novel insight into the individual as well as the collective quantum behavior of matter at ultra-low temperatures. These quantum gases find extensive use in  experiments ranging from studying quantum degeneracy \cite{anderson1995observation}, quantum simulation \cite{sarkar2017nonexponential,hofstetter2018quantum,simon2011quantum} to quantum information processing \cite{garcia2005quantum}, inertial sensing \cite{PhysRevLett.111.083001} and making the most accurate atomic clock \cite{ludlow2015optical,nicholson2015systematic,marti2018imaging}. One of the basic requirements in these experiments is to have a high flux atomic source capable of producing a highly collimated beam of atoms. Such an atomic beam can be slowed down in a collision-free environment using Zeeman slower \cite{PhysRevA.79.063631} and is used for loading into a Magneto-Optic Trap (MOT) for further laser-cooling. Therefore, designing an  efficient atomic oven for the atomic species under consideration \cite{ross1995high} is the first crucial step towards the generation of laser-cooled atomic samples. Atomic species with relatively low vapor pressure (e.g., 1 Pa at 796 K for Strontium atoms) require the oven to be operated at a high temperature (e.g., greater than 700 K for Sr). For such systems, the isolation of the rest of the vacuum chamber from this high-temperature region is technically challenging. This problem is traditionally addressed  either (1) by using a glass or a ceramic break between the oven and the vacuum chamber, Or (2) by incorporating the oven completely inside the Ultra High Vacuum (UHV) chamber \cite{schioppo2012compact,ross1995high}. These measures are good enough for the problem under consideration, but they are relatively complex in terms of design, construction and maintenance. The later solution is quite appealing, however, due to thermal cycling, the heating elements kept inside the vacuum chamber sometimes become fragile and need to be replaced by venting the UHV system, which can be time consuming and tedious. Hence, it is advantageous to keep the heating elements outside the vacuum system. 

In this article, we demonstrate a design that is simple and easy to machine for achieving high-temperatures for an oven of Strontium atoms. This design overcomes the problems mentioned above. Additionally, the oven also removes complications of individually heating the reservoir and collimating capillaries \cite{senaratne2015effusive,song2016cost} and can easily be implemented for other low vapor pressure atomic species. This paper describes a detailed design and construction of such an oven followed by the steady state thermal profile simulation of the system in Finite Element Analysis (FEM) by COMSOL Multiphysics$^{\tiny{\textregistered}}$ software \cite{comsol}. The numerically computed thermal profile has a good agreement to that of experimentally measured values in our system. In the later part of this paper, we have proposed minor changes in the existing design which makes it more suitable in terms of thermal isolation. These modifications are suggested by studying the temperature profile from FEM simulations, using the same module and are more apt for low vapor pressure atomic species. 
\\

\section{Design and Construction}

The oven consists of two sections. The first part is machined from a single block of UHV compatible stainless steel (SS304) and shaped into a CF35 vacuum flange as shown in figure 1. The relevant dimensions of the different parts are shown in figure 1(b). The novelty of this design lies in choosing a geometry that reduces the thermal conductivity between the oven and the rest of the system. For a given material, the thermal conductivity between two points can be reduced by decreasing its thickness and by increasing the distance between them. In the present case, in order to reduce the thermal conductivity, a narrow constriction of thickness 4.5 mm is made by scraping out the material along the radial direction. The length of this narrow neck is limited by the geometry of the current design to 4.5 mm. A hole with diameter 7 mm and length 39 mm is made on this section which serves as a reservoir for 99.9 \% pure Strontium atoms (Sigma Aldrich, 460346-5G). The second part of the oven houses a bundle of narrow capillaries. It has been shown that an array of micro capillary channels outperforms a single aperture collimation tube in terms of intensity and collimation ratio along the axial direction \cite{hanes1960multiple}. If the system is not in molecular flow regime (Knudsen number $<$ 0.5), the collimation is determined by the aspect ratio of the micro-capillary channel. As the aspect ratio is the key to the degree of collimation, for the current experiment, readily available hypodermic needles of 24 Gauge (internal diameter (ID): 400 $\mu$m) are used.
\begin{figure}[ht]
\centering
\includegraphics[width=1\linewidth]{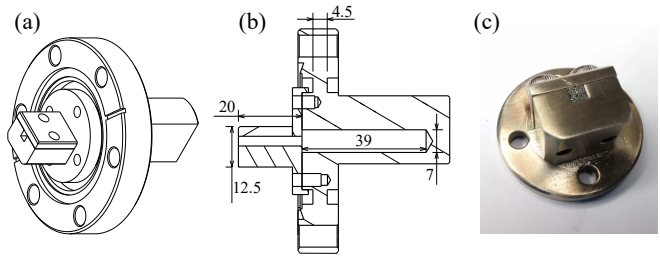} 
\caption{(a) Isometric view of the oven assembly; (b) Schematic diagram of Sr reservoir. All the dimensions are in mm; (c) Image of the needle holder carrying $\sim$60 hypodermic needles. }
\end{figure}
The needles are cut to a length of 15 mm by using a sharp knife. After cutting, the faces are smoothened using fine emery paper. Each needle gives a collimation of $\sim$26 mrad. Approximately 60 needles are bunched together with the help of a stack holder as shown in figure 1(c). An important criteria for smooth operation of such a microchannel-based oven is to make sure that all the capillaries are aligned axially in the direction of atom-capture region. One of the several ways to make a defect-free lattice of such tubes is to use a holder with triangular cross-section \cite{senaratne2015effusive} which results in the hexagonal packing of the needles. For convenience in machining, the design with the square cross-section is chosen in our experiment. The resulting defect in the alignment has not led to any observable problems. The needle holder and the reservoir are assembled with the help of four M4 screws as shown in figure 1(a).

The oven is electrically heated using a 21 AWG (American Wire Gauge) Nichrome wire. A layer of seven turns of such Nichrome wire is wound on the reservoir for this purpose. In order to avoid any electrical short, a  Mica sheet is used between the oven and the heating wire. A cylindrical enclosure (ID 70 mm, length 150 mm) made of Aluminum  is used to isolate the oven region from the surrounding. This cylinder is filled with glass wool to reduce heat loss directly to the environment by convection. The temperature of the oven is maintained by using a switching module (SELEC: TC533). It periodically turns the current ON to maintain a particular set temperature. During each ON time, it passes an alternating current of 8.3 A with a voltage drop of 22.5 V$_{rms}$ controlled by a variac. By changing the duty cycle, it controls the power delivered to the oven and hence maintains the temperature of the oven. The power delivered to the system has been computed by measuring the duty cycle of the switching. The electrical power delivered as a function of oven temperature is shown in figure 2.   

\begin{figure}[ht]
\centering
\includegraphics[width=0.8\linewidth]{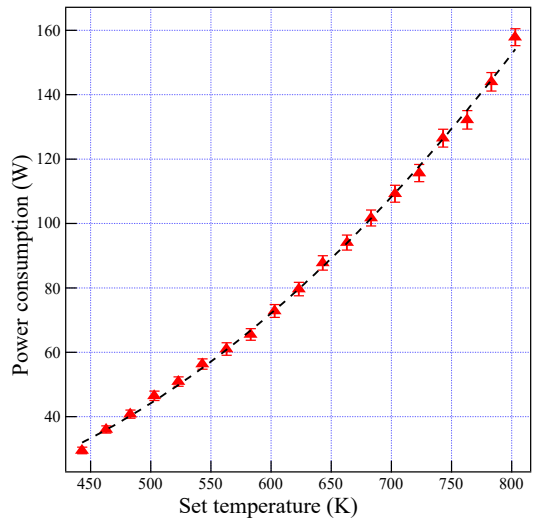} 
\caption{Estimated electrical power as a function of set temperature. Symbols represent experimentally measured data point. The dashed line is a guide to eye. The error bars represent the uncertainty in the estimation of power due to an error in the measurement of duty cycle.}
\end{figure}

\section{Characterization and Results}

In order to understand the temperature distribution of the oven and the effect of modifications in geometry, heat simulations were run in COMSOL Multiphysics (Version-5.2) software. COMSOL uses the finite element method (FEM) to calculate an approximate numerical solution of the required partial differential equation (PDE) for a given set of boundary conditions. The oven attains thermal equilibrium by dissipating heat to the surrounding ambient air via convection and radiation.
To accommodate these factors in the simulation of the temperature profile, we employ the `heat transfer in solid' module in COMSOL. This module allows one to simulate the steady-state/time-dependent solution of the temperature profile in the presence of fluid and solid elements. It utilizes experimentally determined correlations between the temperature profile and fluid flow around a particular geometry to calculate the thermal profile of a given system. The geometry for implementing forced convection in our system is taken to be `planar'. To employ forced convection cooling on a surface we take the characteristic length to be the largest relevant dimension in that region e.g. the characteristc length for the surfaces of ion pump is taken to be 0.3 m. Solutions thus obtained under these conditions are found to converge to reasonable values.

\begin{figure}[ht]
\begin{center}
\includegraphics[width=0.85\linewidth]{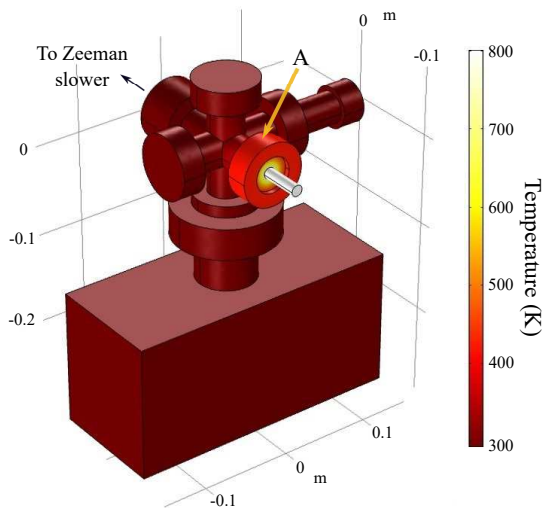} 
\caption{Numerically simulated thermal profile of the oven including a CF35 6-way cross, a 55 $l$/sec ion pump (bottom) and an atomic beam shutter (right). The reservoir is maintained at the constant temperature of 800 K. The temperature is monitored at the region connected to 6-way cross, shown by point `A'.}
\end{center}
\end{figure} 

Figure 3. displays a numerically simulated thermal profile of the system. In our experimental setup, the oven is connected to a CF35 6-way cross, a 55 $l/$s ion pump and a pneumatically operated rotary shutter (Part number - MD20RAIX000Z from UHV Design). A simplified model of the system has been considered for the numerical simulation using COMSOL Multiphysics. The temperature of the reservoir is varied from 500 K to 1300 K in the simulations. For forced convective cooling, the air speed is taken to be 1.5 m/s and the characteristic length at a surface to be the largest local dimension as mentioned previously. Radiative cooling is taken into account by considering the surrounding temperature to be 20$^{\circ}$C and the emissivity of stainless steel (grade SS304) to be 0.65 \cite {zhu2017normal} at all the surfaces under consideration (outside as well as inside the vacuum). After each such simulation, the temperature was measured at the periphery of the oven flange marked as point `A' in figure 3. These simulation parameters give  results comparable to the measured temperature profile at the flange within a reasonable error bound.

\begin{figure}[t]
\begin{center}
   \includegraphics[width=0.9\linewidth]{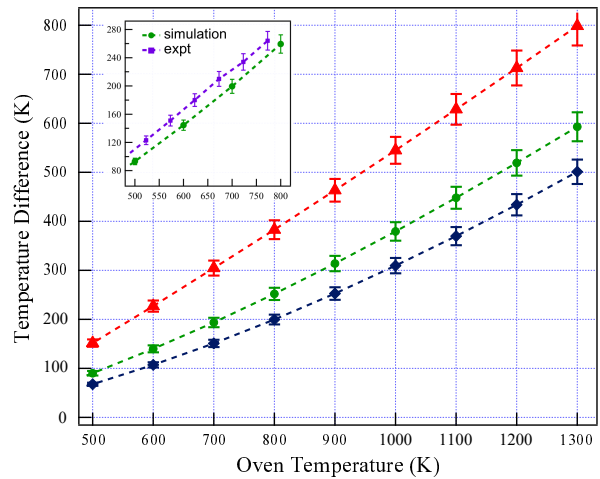} 
   \end{center}
    \caption{Effect of geometry of the constriction on thermal isolation: The numerically simulated temperature difference as a function of oven temperature (a) without any constriction (diamond symbols), (b) with current design of constriction (circular symbols), (c) with proposed design (triangular symbols). Inset shows a comparison between the numerically simulated (circular symbols) and experimentally measured (square symbols) temperature difference as a function of oven temperature. \label{fig:figd} Error bars represent the uncertainty in the values of temperature difference.}
\end{figure}

Figure 4. summarizes the effect of oven geometry on the degree of thermal isolation as measured by the temperature difference between the oven and point `A' on the flange. The numerically simulated values of such temperature difference have been plotted as a function of oven temperature in figure 4.  Simulations were repeated for different oven geometries; (a) without any constriction (diamond symbols), (b) with the current design of the oven \big[length and thickness of the constriction equal to 4.5 mm (circular symbols)\big], and finally (c) with the proposed design of the oven \big[length and thickness of the constriction equal to 13.5 mm and 1.5 mm respectively (triangular symbols)\big] mentioned later in the text. The observed temperature difference between the reservoir and at point `A' of the  flange scales linearly with oven temperature at higher temperatures. In the inset of figure 4, a comparison between the experimental result and numerically simulated values for the current setup are displayed. A $\pm$5\% error has been considered for both the cases to incorporate the uncertainty in the values of emissivity and air current. The experimentally measured thermal profile shows promising agreement with the simulated one. The offset ($\sim$20 K ) between these two curves can be attributed to the fact that the simulated model is not an exact 3-D model. The connections between various flanges in the real system is poorer  as opposed to the simulated system, since in the simulated model, the six way cross together with the flanges is taken as a homogeneous material. We have also omitted the Zeeman slower connected to the opposite side of the Oven.  In the actual system, the various flanges are connected with copper gaskets and stainless steel fasteners. In addition, the emissivity is non uniform since there are slightly different materials near the ion pump and the rotary shutter. 

\begin{figure}[t]
\centering
\includegraphics[width=1\linewidth]{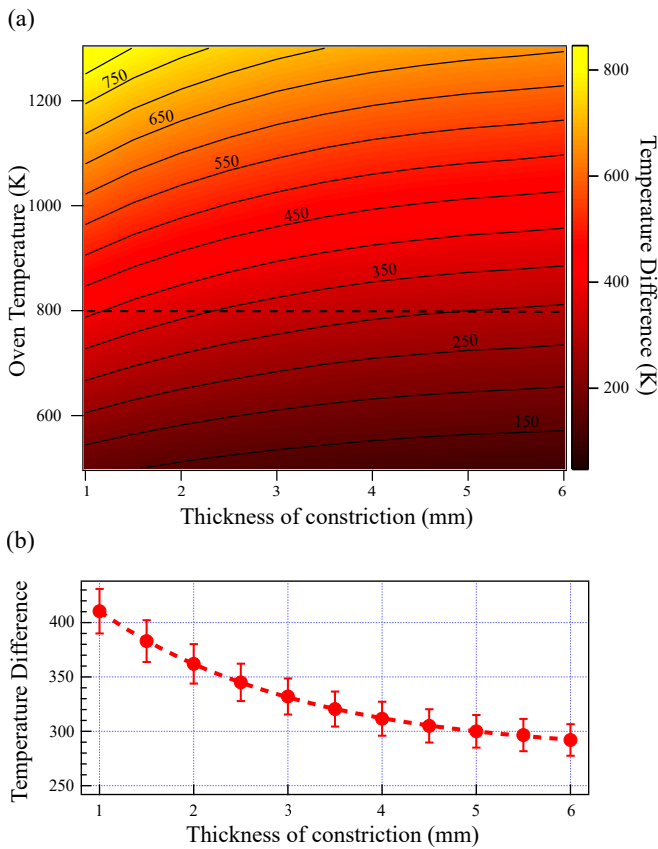} 
\caption{ (a) Contour plot of the difference in temperature as a function of oven temperature and thickness of the constriction (length of the constriction = 13.5 mm), (b) variation of temperature difference as a function of thickness of the constriction for fixed oven temperature at 800 K (Cross section of the contour along the dotted line). Dashed line is a guide to eye.}
\end{figure}

In order to understand the effect of the geometry on the thermal isolation, a series of simulations were done by varying the thickness of the constriction (for a fixed length of 13.5 mm) and the oven temperature.  The results of these simulations are  summarized in figure 5(a) as a contour plot. In order to clearly see the effect of the thickness, a plot of temperature difference as a function of the same is shown in figure 5(b) while keeping the operating temperature of the oven constant at 800 K. The variation of temperature shows power law behavior with  thickness. This leads us to propose a modified design of the oven. In this new design, the length of the narrow part is increased to 13.5 mm, and the thickness of constriction is reduced to 1.5 mm for better thermal resistance, without compromising on the physical strength of the oven. To compare this design with the existing one, the simulation results for this oven design (triangular symbol) are also shown in figure 4 along with the other designs. This design with minimal modification outperforms the existing one by a significant factor making it more suitable for elements that have lower vapor pressure than Strontium and need higher operating temperatures.

In order to characterize the oven, the first stage of $^{88}$Sr MOT  is operated on $5s^2\ ^1S_0\longrightarrow 5s\ 5p\ ^1P_1$ transition ($\lambda = 460.7$ nm, saturation intensity $I_s = 42.7 \ $mW/cm$^2$ ) in a standard $\sigma^{+}-\sigma^{-}$ configuration. The blue light for the experiment is generated using a home built optical frequency doubler by injecting an input light of 922 nm wavelength and power $\sim$700 mW, derived from the output of a commercial tapered amplifier. The frequency of this light is stabilized to an atomic reference\cite{verma2017compact}. Magnetic field for the MOT is generated using a pair of anti-Helmholtz coil generating a magnetic field-gradient of $\sim$12 Gauss/cm/A along the axial direction of the coils. During loading of atoms in the MOT, the quadrupole coils are operated at a current of 4 A and detuning of the MOT laser beams is kept at -33 MHz ($\sim$1 $\Gamma/2\pi$, where $\Gamma$ is the natural linewidth of the cooling transition) from the resonant transition frequency. This leads to decay of a fraction of atoms into the metastable state $ ^3P_2$. In order to bring the atoms back into the main cooling cycle, a repumping laser is operated between the states $5s\ 5p\ ^3P_2\longrightarrow 5s\ 6s\ ^3S_1$ at a wavelength of 707.2 nm. This laser is frequency stabilized using a commercial wavemeter (High Finesse, WSU-30) and a Labview interfaced digital PID. The loading curve of the MOT is obtained by collecting the fluorescence of the trapped atoms onto a Hamamatsu photomultiplier tube (H9307-02) using a lens subtending a solid angle of 0.2 srad at the center of the trap. The effect of repumper is seen in the enhancement of fluorescence by nearly 3 times as shown in figure 5. This however does not prevent the loss of atoms in another triplet state $ ^3P_0$, which requires another repumping laser at 679.2 nm. For the current geometry, we obtain $ 7 \times 10^{5}$ atoms with the loading rate of $2.6\times10^{7}$ atoms/sec.

\vspace{7mm}
\begin{figure}[h]
	\centering
	\includegraphics[width=0.85\linewidth]{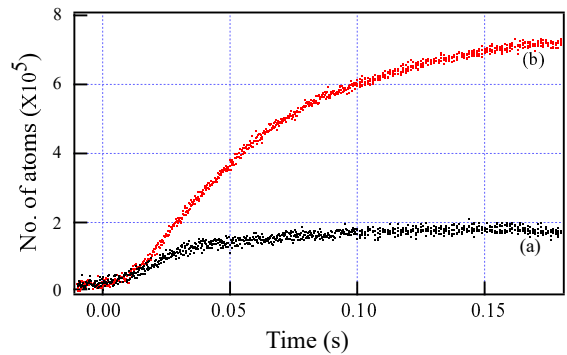} 
	\caption{Fluroscene of Sr atoms in MOT collected on a photo-multiplier tube as a function of time; (a) without repumping laser, (b) with repumper laser.}
\end{figure}

\section{Conclusion}
We have demonstrated a simple, compact, cost-efficient solution for design and construction of  an atomic beam oven for low vapor pressure elements. The ease of machining, freedom of having all the heating elements outside the vacuum chamber makes it more frugal compared to the popular techniques. The measured results of the temperature gradient of the oven agree with  numerical simulations performed using COMSOL Multiphysics. The proposed modification in the design is easy to implement and can be used for other atomic species.
\\

\section{Acknowledgments}
The authors would like to thank the Department of Science and Technology, Govt. of India for grants through EMR/2014/000365 and to IISER Pune for generous funding for this project. We would also like to thank G. V. Pavan Kumar for lending us the COMSOL software package.


\begin{thebibliography}{18}%
\makeatletter
\providecommand \@ifxundefined [1]{%
 \@ifx{#1\undefined}
}%
\providecommand \@ifnum [1]{%
 \ifnum #1\expandafter \@firstoftwo
 \else \expandafter \@secondoftwo
 \fi
}%
\providecommand \@ifx [1]{%
 \ifx #1\expandafter \@firstoftwo
 \else \expandafter \@secondoftwo
 \fi
}%
\providecommand \natexlab [1]{#1}%
\providecommand \enquote  [1]{``#1''}%
\providecommand \bibnamefont  [1]{#1}%
\providecommand \bibfnamefont [1]{#1}%
\providecommand \citenamefont [1]{#1}%
\providecommand \href@noop [0]{\@secondoftwo}%
\providecommand \href [0]{\begingroup \@sanitize@url \@href}%
\providecommand \@href[1]{\@@startlink{#1}\@@href}%
\providecommand \@@href[1]{\endgroup#1\@@endlink}%
\providecommand \@sanitize@url [0]{\catcode `\\12\catcode `\$12\catcode
  `\&12\catcode `\#12\catcode `\^12\catcode `\_12\catcode `\%12\relax}%
\providecommand \@@startlink[1]{}%
\providecommand \@@endlink[0]{}%
\providecommand \url  [0]{\begingroup\@sanitize@url \@url }%
\providecommand \@url [1]{\endgroup\@href {#1}{\urlprefix }}%
\providecommand \urlprefix  [0]{URL }%
\providecommand \Eprint [0]{\href }%
\providecommand \doibase [0]{http://dx.doi.org/}%
\providecommand \selectlanguage [0]{\@gobble}%
\providecommand \bibinfo  [0]{\@secondoftwo}%
\providecommand \bibfield  [0]{\@secondoftwo}%
\providecommand \translation [1]{[#1]}%
\providecommand \BibitemOpen [0]{}%
\providecommand \bibitemStop [0]{}%
\providecommand \bibitemNoStop [0]{.\EOS\space}%
\providecommand \EOS [0]{\spacefactor3000\relax}%
\providecommand \BibitemShut  [1]{\csname bibitem#1\endcsname}%
\let\auto@bib@innerbib\@empty
\bibitem [{\citenamefont {Anderson}\ \emph {et~al.}(1995)\citenamefont
  {Anderson}, \citenamefont {Ensher}, \citenamefont {Matthews}, \citenamefont
  {Wieman},\ and\ \citenamefont {Cornell}}]{anderson1995observation}%
  \BibitemOpen
  \bibfield  {author} {\bibinfo {author} {\bibfnamefont {M.~H.}\ \bibnamefont
  {Anderson}}, \bibinfo {author} {\bibfnamefont {J.~R.}\ \bibnamefont
  {Ensher}}, \bibinfo {author} {\bibfnamefont {M.~R.}\ \bibnamefont
  {Matthews}}, \bibinfo {author} {\bibfnamefont {C.~E.}\ \bibnamefont
  {Wieman}}, \ and\ \bibinfo {author} {\bibfnamefont {E.~A.}\ \bibnamefont
  {Cornell}},\ }\bibfield  {title} {\enquote {\bibinfo {title} {Observation of
  bose-einstein condensation in a dilute atomic vapor},}\ }\href {\doibase
  10.1126/science.269.5221.198} {\bibfield  {journal} {\bibinfo  {journal}
  {Science}\ }\textbf {\bibinfo {volume} {269}},\ \bibinfo {pages} {198--201}
  (\bibinfo {year} {1995})}\BibitemShut {NoStop}%
\bibitem [{\citenamefont {Sarkar}\ \emph {et~al.}(2017)\citenamefont {Sarkar},
  \citenamefont {Paul}, \citenamefont {Vishwakarma}, \citenamefont {Kumar},
  \citenamefont {Verma}, \citenamefont {Sainath}, \citenamefont {Rapol},\ and\
  \citenamefont {Santhanam}}]{sarkar2017nonexponential}%
  \BibitemOpen
  \bibfield  {author} {\bibinfo {author} {\bibfnamefont {S.}~\bibnamefont
  {Sarkar}}, \bibinfo {author} {\bibfnamefont {S.}~\bibnamefont {Paul}},
  \bibinfo {author} {\bibfnamefont {C.}~\bibnamefont {Vishwakarma}}, \bibinfo
  {author} {\bibfnamefont {S.}~\bibnamefont {Kumar}}, \bibinfo {author}
  {\bibfnamefont {G.}~\bibnamefont {Verma}}, \bibinfo {author} {\bibfnamefont
  {M.}~\bibnamefont {Sainath}}, \bibinfo {author} {\bibfnamefont {U.~D.}\
  \bibnamefont {Rapol}}, \ and\ \bibinfo {author} {\bibfnamefont {M.~S.}\
  \bibnamefont {Santhanam}},\ }\bibfield  {title} {\enquote {\bibinfo {title}
  {Nonexponential decoherence and subdiffusion in atom-optics kicked rotor},}\
  }\href {\doibase 10.1103/PhysRevLett.118.174101} {\bibfield  {journal}
  {\bibinfo  {journal} {Phys. Rev. Lett.}\ }\textbf {\bibinfo {volume} {118}},\
  \bibinfo {pages} {174101} (\bibinfo {year} {2017})}\BibitemShut {NoStop}%
\bibitem [{\citenamefont {Hofstetter}\ and\ \citenamefont
  {Qin}(2018)}]{hofstetter2018quantum}%
  \BibitemOpen
  \bibfield  {author} {\bibinfo {author} {\bibfnamefont {W.}~\bibnamefont
  {Hofstetter}}\ and\ \bibinfo {author} {\bibfnamefont {T.}~\bibnamefont
  {Qin}},\ }\bibfield  {title} {\enquote {\bibinfo {title} {Quantum simulation
  of strongly correlated condensed matter systems},}\ }\href
  {http://stacks.iop.org/0953-4075/51/i=8/a=082001} {\bibfield  {journal}
  {\bibinfo  {journal} {J. Phys. B: At. Mol. and Opt.
  Phys.}\ }\textbf {\bibinfo {volume} {51}},\ \bibinfo {pages} {082001}
  (\bibinfo {year} {2018})}\BibitemShut {NoStop}%
\bibitem [{\citenamefont {Simon}\ \emph {et~al.}(2011)\citenamefont {Simon},
  \citenamefont {Bakr}, \citenamefont {Ma}, \citenamefont {Tai}, \citenamefont
  {Preiss},\ and\ \citenamefont {Greiner}}]{simon2011quantum}%
  \BibitemOpen
  \bibfield  {author} {\bibinfo {author} {\bibfnamefont {J.}~\bibnamefont
  {Simon}}, \bibinfo {author} {\bibfnamefont {W.~S.}\ \bibnamefont {Bakr}},
  \bibinfo {author} {\bibfnamefont {R.}~\bibnamefont {Ma}}, \bibinfo {author}
  {\bibfnamefont {M.~E.}\ \bibnamefont {Tai}}, \bibinfo {author} {\bibfnamefont
  {P.~M.}\ \bibnamefont {Preiss}}, \ and\ \bibinfo {author} {\bibfnamefont
  {M.}~\bibnamefont {Greiner}},\ }\bibfield  {title} {\enquote {\bibinfo
  {title} {Quantum simulation of antiferromagnetic spin chains in an optical
  lattice},}\ }\href {http://dx.doi.org/10.1038/nature09994} {\bibfield
  {journal} {\bibinfo  {journal} {Nature}\ }\textbf {\bibinfo {volume} {472}},\
  \bibinfo {pages} {307} (\bibinfo {year} {2011})}\BibitemShut {NoStop}%
\bibitem [{\citenamefont {García-Ripoll}, \citenamefont {Zoller},\ and\
  \citenamefont {Cirac}(2005)}]{garcia2005quantum}%
  \BibitemOpen
  \bibfield  {author} {\bibinfo {author} {\bibfnamefont {J.~J.}\ \bibnamefont
  {García-Ripoll}}, \bibinfo {author} {\bibfnamefont {P.}~\bibnamefont
  {Zoller}}, \ and\ \bibinfo {author} {\bibfnamefont {J.~I.}\ \bibnamefont
  {Cirac}},\ }\bibfield  {title} {\enquote {\bibinfo {title} {Quantum
  information processing with cold atoms and trapped ions},}\ }\href
  {http://stacks.iop.org/0953-4075/38/i=9/a=008} {\bibfield  {journal}
  {\bibinfo  {journal} {J. Phys. B:  At. Mol. and Opt.
  Phys.}\ }\textbf {\bibinfo {volume} {38}},\ \bibinfo {pages} {S567}
  (\bibinfo {year} {2005})}\BibitemShut {NoStop}%
\bibitem [{\citenamefont {Dickerson}\ \emph {et~al.}(2013)\citenamefont
  {Dickerson}, \citenamefont {Hogan}, \citenamefont {Sugarbaker}, \citenamefont
  {Johnson},\ and\ \citenamefont {Kasevich}}]{PhysRevLett.111.083001}%
  \BibitemOpen
  \bibfield  {author} {\bibinfo {author} {\bibfnamefont {S.~M.}\ \bibnamefont
  {Dickerson}}, \bibinfo {author} {\bibfnamefont {J.~M.}\ \bibnamefont
  {Hogan}}, \bibinfo {author} {\bibfnamefont {A.}~\bibnamefont {Sugarbaker}},
  \bibinfo {author} {\bibfnamefont {D.~M.~S.}\ \bibnamefont {Johnson}}, \ and\
  \bibinfo {author} {\bibfnamefont {M.~A.}\ \bibnamefont {Kasevich}},\
  }\bibfield  {title} {\enquote {\bibinfo {title} {Multiaxis inertial sensing
  with long-time point source atom interferometry},}\ }\href {\doibase
  10.1103/PhysRevLett.111.083001} {\bibfield  {journal} {\bibinfo  {journal}
  {Phys. Rev. Lett.}\ }\textbf {\bibinfo {volume} {111}},\ \bibinfo {pages}
  {083001} (\bibinfo {year} {2013})}\BibitemShut {NoStop}%
\bibitem [{\citenamefont {Ludlow}\ \emph {et~al.}(2015)\citenamefont {Ludlow},
  \citenamefont {Boyd}, \citenamefont {Ye}, \citenamefont {Peik},\ and\
  \citenamefont {Schmidt}}]{ludlow2015optical}%
  \BibitemOpen
  \bibfield  {author} {\bibinfo {author} {\bibfnamefont {A.~D.}\ \bibnamefont
  {Ludlow}}, \bibinfo {author} {\bibfnamefont {M.~M.}\ \bibnamefont {Boyd}},
  \bibinfo {author} {\bibfnamefont {J.}~\bibnamefont {Ye}}, \bibinfo {author}
  {\bibfnamefont {E.}~\bibnamefont {Peik}}, \ and\ \bibinfo {author}
  {\bibfnamefont {P.~O.}\ \bibnamefont {Schmidt}},\ }\bibfield  {title}
  {\enquote {\bibinfo {title} {Optical atomic clocks},}\ }\href {\doibase
  10.1103/RevModPhys.87.637} {\bibfield  {journal} {\bibinfo  {journal} {Rev.
  Mod. Phys.}\ }\textbf {\bibinfo {volume} {87}},\ \bibinfo {pages} {637--701}
  (\bibinfo {year} {2015})}\BibitemShut {NoStop}%
\bibitem [{\citenamefont {Nicholson}\ \emph {et~al.}(2015)\citenamefont
  {Nicholson}, \citenamefont {Campbell}, \citenamefont {Hutson}, \citenamefont
  {Marti}, \citenamefont {Bloom}, \citenamefont {McNally}, \citenamefont
  {Zhang}, \citenamefont {Barrett}, \citenamefont {Safronova}, \citenamefont
  {Strouse} \emph {et~al.}}]{nicholson2015systematic}%
  \BibitemOpen
  \bibfield  {author} {\bibinfo {author} {\bibfnamefont {T.}~\bibnamefont
  {Nicholson}}, \bibinfo {author} {\bibfnamefont {S.}~\bibnamefont {Campbell}},
  \bibinfo {author} {\bibfnamefont {R.}~\bibnamefont {Hutson}}, \bibinfo
  {author} {\bibfnamefont {G.}~\bibnamefont {Marti}}, \bibinfo {author}
  {\bibfnamefont {B.}~\bibnamefont {Bloom}}, \bibinfo {author} {\bibfnamefont
  {R.}~\bibnamefont {McNally}}, \bibinfo {author} {\bibfnamefont
  {W.}~\bibnamefont {Zhang}}, \bibinfo {author} {\bibfnamefont
  {M.}~\bibnamefont {Barrett}}, \bibinfo {author} {\bibfnamefont
  {M.}~\bibnamefont {Safronova}}, \bibinfo {author} {\bibfnamefont
  {G.}~\bibnamefont {Strouse}},  \emph {et~al.},\ }\bibfield  {title} {\enquote
  {\bibinfo {title} {Systematic evaluation of an atomic clock at 2$\times$ 10$^{-
  18}$ total uncertainty},}\ }\href {http://dx.doi.org/10.1038/ncomms7896}
  {\bibfield  {journal} {\bibinfo  {journal} {Nat. Commun.}\ }\textbf
  {\bibinfo {volume} {6}},\ \bibinfo {pages} {6896} (\bibinfo {year}
  {2015})}\BibitemShut {NoStop}%
\bibitem [{\citenamefont {Marti}\ \emph {et~al.}(2018)\citenamefont {Marti},
  \citenamefont {Hutson}, \citenamefont {Goban}, \citenamefont {Campbell},
  \citenamefont {Poli},\ and\ \citenamefont {Ye}}]{marti2018imaging}%
  \BibitemOpen
  \bibfield  {author} {\bibinfo {author} {\bibfnamefont {G.~E.}\ \bibnamefont
  {Marti}}, \bibinfo {author} {\bibfnamefont {R.~B.}\ \bibnamefont {Hutson}},
  \bibinfo {author} {\bibfnamefont {A.}~\bibnamefont {Goban}}, \bibinfo
  {author} {\bibfnamefont {S.~L.}\ \bibnamefont {Campbell}}, \bibinfo {author}
  {\bibfnamefont {N.}~\bibnamefont {Poli}}, \ and\ \bibinfo {author}
  {\bibfnamefont {J.}~\bibnamefont {Ye}},\ }\bibfield  {title} {\enquote
  {\bibinfo {title} {Imaging optical frequencies with $100\text{ }\text{
  }\ensuremath{\mu}\mathrm{Hz}$ precision and $1.1\text{ }\text{
  }\ensuremath{\mu}\mathrm{m}$ resolution},}\ }\href {\doibase
  10.1103/PhysRevLett.120.103201} {\bibfield  {journal} {\bibinfo  {journal}
  {Phys. Rev. Lett.}\ }\textbf {\bibinfo {volume} {120}},\ \bibinfo {pages}
  {103201} (\bibinfo {year} {2018})}\BibitemShut {NoStop}%
\bibitem [{\citenamefont {Lin}\ \emph {et~al.}(2009)\citenamefont {Lin},
  \citenamefont {Perry}, \citenamefont {Compton}, \citenamefont {Spielman},\
  and\ \citenamefont {Porto}}]{PhysRevA.79.063631}%
  \BibitemOpen
  \bibfield  {author} {\bibinfo {author} {\bibfnamefont {Y.-J.}\ \bibnamefont
  {Lin}}, \bibinfo {author} {\bibfnamefont {A.~R.}\ \bibnamefont {Perry}},
  \bibinfo {author} {\bibfnamefont {R.~L.}\ \bibnamefont {Compton}}, \bibinfo
  {author} {\bibfnamefont {I.~B.}\ \bibnamefont {Spielman}}, \ and\ \bibinfo
  {author} {\bibfnamefont {J.~V.}\ \bibnamefont {Porto}},\ }\bibfield  {title}
  {\enquote {\bibinfo {title} {Rapid production of $^{87}\text{R}\text{b}$
  bose-einstein condensates in a combined magnetic and optical potential},}\
  }\href {\doibase 10.1103/PhysRevA.79.063631} {\bibfield  {journal} {\bibinfo
  {journal} {Phys. Rev. A}\ }\textbf {\bibinfo {volume} {79}},\ \bibinfo
  {pages} {063631} (\bibinfo {year} {2009})}\BibitemShut {NoStop}%
\bibitem [{\citenamefont {Ross}\ and\ \citenamefont
  {Sonntag}(1995)}]{ross1995high}%
  \BibitemOpen
  \bibfield  {author} {\bibinfo {author} {\bibfnamefont {K.~J.}\ \bibnamefont
  {Ross}}\ and\ \bibinfo {author} {\bibfnamefont {B.}~\bibnamefont {Sonntag}},\
  }\bibfield  {title} {\enquote {\bibinfo {title} {High temperature metal atom
  beam sources},}\ }\href {\doibase 10.1063/1.1145337} {\bibfield  {journal}
  {\bibinfo  {journal} {Rev. Sci. Instrum.}\ }\textbf {\bibinfo
  {volume} {66}},\ \bibinfo {pages} {4409--4433} (\bibinfo {year}
  {1995})}\BibitemShut {NoStop}%
\bibitem [{\citenamefont {Schioppo}\ \emph {et~al.}(2012)\citenamefont
  {Schioppo}, \citenamefont {Poli}, \citenamefont {Prevedelli}, \citenamefont
  {Falke}, \citenamefont {Lisdat}, \citenamefont {Sterr},\ and\ \citenamefont
  {Tino}}]{schioppo2012compact}%
  \BibitemOpen
  \bibfield  {author} {\bibinfo {author} {\bibfnamefont {M.}~\bibnamefont
  {Schioppo}}, \bibinfo {author} {\bibfnamefont {N.}~\bibnamefont {Poli}},
  \bibinfo {author} {\bibfnamefont {M.}~\bibnamefont {Prevedelli}}, \bibinfo
  {author} {\bibfnamefont {S.}~\bibnamefont {Falke}}, \bibinfo {author}
  {\bibfnamefont {C.}~\bibnamefont {Lisdat}}, \bibinfo {author} {\bibfnamefont
  {U.}~\bibnamefont {Sterr}}, \ and\ \bibinfo {author} {\bibfnamefont {G.~M.}\
  \bibnamefont {Tino}},\ }\bibfield  {title} {\enquote {\bibinfo {title} {A
  compact and efficient strontium oven for laser-cooling experiments},}\ }\href
  {\doibase 10.1063/1.4756936} {\bibfield  {journal} {\bibinfo  {journal}
  {Rev. Sci. Instrum.}\ }\textbf {\bibinfo {volume} {83}},\
  \bibinfo {pages} {103101} (\bibinfo {year} {2012})}\BibitemShut {NoStop}%
\bibitem [{\citenamefont {Senaratne}\ \emph {et~al.}(2015)\citenamefont
  {Senaratne}, \citenamefont {Rajagopal}, \citenamefont {Geiger}, \citenamefont
  {Fujiwara}, \citenamefont {Lebedev},\ and\ \citenamefont
  {Weld}}]{senaratne2015effusive}%
  \BibitemOpen
  \bibfield  {author} {\bibinfo {author} {\bibfnamefont {R.}~\bibnamefont
  {Senaratne}}, \bibinfo {author} {\bibfnamefont {S.~V.}\ \bibnamefont
  {Rajagopal}}, \bibinfo {author} {\bibfnamefont {Z.~A.}\ \bibnamefont
  {Geiger}}, \bibinfo {author} {\bibfnamefont {K.~M.}\ \bibnamefont
  {Fujiwara}}, \bibinfo {author} {\bibfnamefont {V.}~\bibnamefont {Lebedev}}, \
  and\ \bibinfo {author} {\bibfnamefont {D.~M.}\ \bibnamefont {Weld}},\
  }\bibfield  {title} {\enquote {\bibinfo {title} {Effusive atomic oven nozzle
  design using an aligned microcapillary array},}\ }\href {\doibase
  10.1063/1.4907401} {\bibfield  {journal} {\bibinfo  {journal} {Rev. Sci. Instrum.}\ }\textbf {\bibinfo {volume} {86}},\ \bibinfo {pages}
  {023105} (\bibinfo {year} {2015})}\BibitemShut {NoStop}%
\bibitem [{\citenamefont {Song}\ \emph {et~al.}(2016)\citenamefont {Song},
  \citenamefont {Zou}, \citenamefont {Zhang}, \citenamefont {Cho},\ and\
  \citenamefont {Jo}}]{song2016cost}%
  \BibitemOpen
  \bibfield  {author} {\bibinfo {author} {\bibfnamefont {B.}~\bibnamefont
  {Song}}, \bibinfo {author} {\bibfnamefont {Y.}~\bibnamefont {Zou}}, \bibinfo
  {author} {\bibfnamefont {S.}~\bibnamefont {Zhang}}, \bibinfo {author}
  {\bibfnamefont {C.-w.}\ \bibnamefont {Cho}}, \ and\ \bibinfo {author}
  {\bibfnamefont {G.-B.}\ \bibnamefont {Jo}},\ }\bibfield  {title} {\enquote
  {\bibinfo {title} {A cost-effective high-flux source of cold ytterbium
  atoms},}\ }\href {\doibase 10.1007/s00340-016-6523-8} {\bibfield  {journal}
  {\bibinfo  {journal} {Appl. Phys. B}\ }\textbf {\bibinfo {volume}
  {122}},\ \bibinfo {pages} {250} (\bibinfo {year} {2016})}\BibitemShut
  {NoStop}%
\bibitem [{\citenamefont {Multiphysics Reference~Manual}()}]{comsol}%
  \BibitemOpen
  \bibfield  {author} {\bibinfo {author} {\bibfnamefont {C.}~\bibnamefont
  {Multiphysics Reference~Manual}},\ }\href@noop {} {\bibfield  {journal}
  {\bibinfo  {journal} {Version 5.2}\ }}\Eprint
  {http://arxiv.org/abs/www.comsol.com} {www.comsol.com} \BibitemShut {NoStop}%
\bibitem [{\citenamefont {Hanes}(1960)}]{hanes1960multiple}%
  \BibitemOpen
  \bibfield  {author} {\bibinfo {author} {\bibfnamefont {G.~R.}\ \bibnamefont
  {Hanes}},\ }\bibfield  {title} {\enquote {\bibinfo {title} {Multiple tube
  collimator for gas beams},}\ }\href {\doibase 10.1063/1.1735519} {\bibfield
  {journal} {\bibinfo  {journal} {J. Appl. Phys.}\ }\textbf
  {\bibinfo {volume} {31}},\ \bibinfo {pages} {2171--2175} (\bibinfo {year}
  {1960})}\BibitemShut {NoStop}%
\bibitem [{\citenamefont {Zhu}\ \emph {et~al.}(2017)\citenamefont {Zhu},
  \citenamefont {Shi}, \citenamefont {Zhu},\ and\ \citenamefont
  {Sun}}]{zhu2017normal}%
  \BibitemOpen
  \bibfield  {author} {\bibinfo {author} {\bibfnamefont {W.}~\bibnamefont
  {Zhu}}, \bibinfo {author} {\bibfnamefont {D.}~\bibnamefont {Shi}}, \bibinfo
  {author} {\bibfnamefont {Z.}~\bibnamefont {Zhu}}, \ and\ \bibinfo {author}
  {\bibfnamefont {J.}~\bibnamefont {Sun}},\ }\bibfield  {title} {\enquote
  {\bibinfo {title} {Normal spectral emissivity models of steel 304 at
  800--1100 K with an oxide layer on the specimen surface},}\ }\href {\doibase
  10.1007/s12666-016-0907-7} {\bibfield  {journal} {\bibinfo  {journal}
  {Transactions of the Indian Institute of Metals}\ }\textbf {\bibinfo {volume}
  {70}},\ \bibinfo {pages} {1083--1090} (\bibinfo {year} {2017})}\BibitemShut
  {NoStop}%
\bibitem [{\citenamefont {Verma}\ \emph {et~al.}(2017)\citenamefont {Verma},
  \citenamefont {Vishwakarma}, \citenamefont {Dharmadhikari},\ and\
  \citenamefont {Rapol}}]{verma2017compact}%
  \BibitemOpen
  \bibfield  {author} {\bibinfo {author} {\bibfnamefont {G.}~\bibnamefont
  {Verma}}, \bibinfo {author} {\bibfnamefont {C.}~\bibnamefont {Vishwakarma}},
  \bibinfo {author} {\bibfnamefont {C.~V.}\ \bibnamefont {Dharmadhikari}}, \
  and\ \bibinfo {author} {\bibfnamefont {U.~D.}\ \bibnamefont {Rapol}},\
  }\bibfield  {title} {\enquote {\bibinfo {title} {A compact atomic beam based
  system for doppler-free laser spectroscopy of strontium atoms},}\ }\href
  {\doibase 10.1063/1.4977593} {\bibfield  {journal} {\bibinfo  {journal}
  {Rev. Sci. Instrum.}\ }\textbf {\bibinfo {volume} {88}},\
  \bibinfo {pages} {033103} (\bibinfo {year} {2017})}\BibitemShut {NoStop}%
\end{thebibliography}
%

\end{document}